\documentclass[aps,prb,twocolumn,groupedaddress]{revtex4-2}
\usepackage[english]{babel}
\usepackage{bm}
\usepackage{amsmath}
\usepackage[dvips]{graphicx}
\usepackage[colorlinks=true, allcolors=blue]{hyperref}
\usepackage{diagbox}

\begin{document}
\title{Plasma solitons in gated two-dimensional electron systems: exactly solvable analytical model for the regime beyond weak non-linearity}

\author{A.~A.~Zabolotnykh}
\affiliation{Kotelnikov Institute of Radio-engineering and Electronics of the RAS, Mokhovaya 11-7, Moscow 125009, Russia}

\begin{abstract}
We analytically study plasma solitary waves, or solitons, in a two-dimensional (2D) electron system (ES) placed in close proximity to and between two ideal metallic gates. As a rule, solitons are described using a perturbative approach applicable only in the weak non-linearity regime. In contrast, we analyze solitons considering a non-perturbative model. This framework enables an exact analytical description of the soliton shape. Moreover, it can be achieved in the regime beyond weak non-linearity --- when the concentration deviation due to the soliton is of the order of the equilibrium concentration. We determine the conditions required for a soliton to exist and derive the relationship between its amplitude, width, and velocity. We believe that our results obtained for the given model can provide valuable insight into the physics of non-linear waves.  
\end{abstract}
\maketitle

Studies of low-dimensional structures are at the heart of modern condensed matter physics. A significant part of the research is devoted to the electronic properties of matter, in particular, the collective charge density excitations --- plasma waves, or plasmons. Plasmons are especially intriguing when considered in so-called gated two-dimensional (2D) electron systems (ESs), with a planar metal electrode (gate) in proximity and parallel to the 2DES. Remarkably, these gated plasmons make possible the detection of terahertz radiation~\cite{Dyakonov1993,Satou2003,Shaner2005,Dyakonov2005,Knap2009,Muravev2012,Lusakowski2016,Bandurin2018}, compression of light~\cite{Iranzo2018,Koppens2020}, study of relativistic effects~\cite{Andreev2021}, etc.

\begin{figure}[t]
\includegraphics[width=1.0\columnwidth]{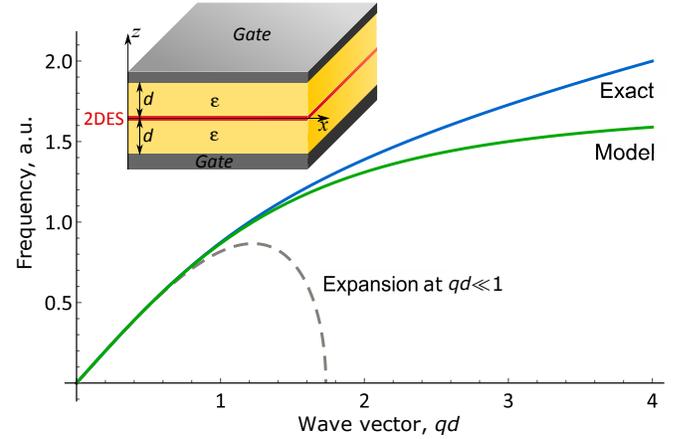}
    \caption{\label{fig:system} The exact spectrum of linear plasmons~(\ref{Eq:spectrum}) (blue) compared against the approximations (gray and green) obtained, respectively, based on the expansion at $qd\ll 1$, and the model Green's function of the Poisson equation~(\ref{Eq:model}). The inset shows the schematic view of the 2DES setup under consideration.
    }
\end{figure}

This Letter presents an analytical investigation of a particular type of non-linear electron-density waves in a gated 2DES --- the solitary plasma waves, or solitons. Solitons are space-localized waves that preserve their shape as they propagate. Qualitatively, the wave-profile stability can be explained by the fact that the non-linearity of the wave propagation equations is compensated for by the dispersion effect, namely, by the dependence of phase velocity on the wave vector (in linear regime). To begin with, let us consider the dispersion law of gated plasmons. To be specific, we examine a 2DES screened by two ideal metal gates positioned above and below it, as illustrated on the inset of Fig.~\ref{fig:system}. Then, the plasmon spectrum can be expressed as follows~\cite{Dahl1977,Ando1982} (we use CGS system of units throughout the Letter):
\begin{equation}
\label{Eq:spectrum}
    \omega^2=\frac{2\pi e^2 n_0}{m\varepsilon}q \tanh{qd}, 
\end{equation}
where $\omega$ and $q$ are, respectively, the plasmon frequency and wave vector in the 2DES plane, $n_0$ is the equilibrium concentration of electrons, $-e$ and $m$ are the electron charge and effective mass, $d$ and $\varepsilon$ are the distance and dielectric permittivity between the 2DES and the gate. We note that in deriving~(\ref{Eq:spectrum}), the 2DES is assumed to be ''clean'', with infinitely large electron relaxation time, and the  electromagnetic retardation effects are neglected. The resultant spectrum is plotted in Fig.~\ref{fig:system} in the solid blue line.

Unlike the 3DES, the complexity of the analytical description of plasmons in a 2DES is associated with solving the Poisson equation, which provides the relation between the potential and charge density in a plasma wave. The challenge arises from the non-local (integral) nature of the relationship since the charges are confined to the 2DES plane, whereas the potential occupies the whole 3D space.
For example, to find the spectrum for linear waves~(\ref{Eq:spectrum}), the Poisson equation is solved first using the Fourier transform in the coordinates along the 2DES plane. Then, the solution to the resultant differential equation is obtained in the $z$-coordinate --- in the direction perpendicular to the 2DES plane (see the inset in Fig.~\ref{fig:system}).
In the case of non-linear waves, however, the given technique becomes inefficient. Hence, describing these waves requires solving a non-linear integral equation, with the exact solution almost impossible to find.
Therefore, in practice, obtaining an analytical description of non-linear waves in low-dimensional systems generally involves an expansion of the Poisson equation into a series governed by a small parameter~\cite{Govorov1999,Narahara2008,Svintsov2013,Koshelev2017,Zdyrski2019}.

Let us illustrate the expansion procedure on the example of plasmons in a gated 2DES. Considering a $\delta$-thin in $z$-direction 2DES, we can use the Poisson equation to express
the relation between the deviation of the 2D electron concentration $n(x,t)$ from its equilibrium value and the potential $\varphi(x,t)$ in the 2DES plane, at $z=0$, as follows:
\begin{equation}
\label{Eq:Po}
    \varphi(x,t)=\frac{-e}{\varepsilon}\int^{+\infty}_{-\infty}G(x-x')n(x',t)dx',
\end{equation}
where the Green's function of the Poisson equation $G(x)$ and its Fourier transform $G(q)$ are defined as
\begin{equation}
    G(x)=-2\ln\tanh\left(\frac{\pi|x|}{4d}\right), \quad G(q)=\frac{2\pi}{q} \tanh(qd),
\end{equation}
and the spectrum in~(\ref{Eq:spectrum}) is related to $G(q)$ as $\omega^2=e^2n_0 q^2 G(q)/m\varepsilon$.
Then, the above-mentioned simplification of integral equation~(\ref{Eq:Po}) is achieved by expanding $G(q)$ into a series with respect to the small parameter $qd\ll 1$.
Thus, the Green's function $G(q)$ is replaced by its series expansion: $G_e(q)= 2\pi d(1-d^2q^2/3)$. Consequently, in $x$-space, $G_e(q)$ becomes $G_e (x) =2\pi d\left(\delta(x)+\delta''(x)d^2/3\right)$, and the Poisson equation (\ref{Eq:Po}) takes the form of a simple differential equation 
\footnote{For the explicit expression of expanded Poisson equation, see Eq.~(2) in Ref.~\cite{Govorov1999}, Eq.~(4) in Ref.~\cite{Narahara2008}, Eq.~(22) in Ref.~\cite{Svintsov2013}, Eq.~(4) in Ref.~\cite{Koshelev2017}, Eq.~(25) in Ref.~\cite{Zdyrski2019}.}.
The plasmon spectrum corresponding to $G_e(q)$ takes the form $\omega^2=2\pi e^2n_0 q^2 d(1-d^2q^2/3)/m\varepsilon$ (gray dashed line in Fig.~\ref{fig:system}). Importantly, since the expansion is valid only for $qd\ll 1$, this approach can be applied to non-linear waves and solitons, in particular, only in the regime of \textit{weak} non-linearity, when the deviation of the electron concentration associated with the soliton $n(x,t)$ is small compared to the equilibrium concentration $n_0$: $|n(x,t)|\ll n_0$
\footnote{for additional discussion on the given issue, see paragraphs prior to and following Eq.~(5) in Ref.~\cite{Govorov1999}}.

In this Letter we analyze solitons in a gated 2DES using a different method that allows to obtain an exact analytical solution with \textit{no} restrictions of weak non-linearities $|n|\ll n_0$ and $qd \ll 1$, unlike the conventional approach of expanding $G(q)$ into the $q$-series. We replace $G(q)$ with the model function $G_{m}(q)$. This function has two key properties: it is proportional to $1/(\alpha^2+q^2d^2)$, where $\alpha$ is the model constant, i.e. it has a simple pole; and it has the same coefficients for the $q^0$ and $q^2$ terms in the $q$-series as $G(q)$~\cite{CarrierBook}. 
The function $G_{m}$ that meets the desired criteria can be formulated as follows:
\begin{equation}
\label{Eq:model}
    G_{m}(x)=\pi \alpha \exp\left(\frac{-\alpha |x|}{d}\right), \, G_{m}(q)= \frac{2\pi d}{1+q^2d^2/\alpha^2},
\end{equation}
and $\alpha=\sqrt{3}$~\footnote{On the whole, the model function $G_m$ can be chosen in different ways. Usually, $G_m(x)$ is taken in the exponential form $A\exp(-B|x|)$, with $A$ and $B>0$ as model constants. This particular choice highly simplifies the Poisson equation and often enables its analytical solution. To find the amplitude constant $A$ and the constant $B$, which corresponds to the inverse
“interaction length”, we use the following two conditions. The first condition is $G(q=0)=G_m(q=0)$. In $x$-space, it corresponds to $\int^{+\infty}_{+\infty} (G(x)-G_m(x))dx=0$, which qualitatively signifies the equality of the ''average interaction potentials'' in the exact and model approaches. The second condition can be determined either by equating the coefficients of $q^2$ terms in the $q$-series for $G(q)$ and $G_m(q)$ \cite{CarrierBook}, or from the same asymptotic behavior of $G(x)$ and $G_m(x)$ at $x\to\infty$ \cite{WhithamBook}. The values of $\alpha$~(\ref{Eq:model}) obtained in these two cases are found to be relatively close, at $\sqrt{3}\approx 1.73$ and $\pi/2\approx 1.57$, respectively.}.

The resultant plasmon spectrum, defined as $\omega^2=2\pi e^2n_0 q^2 d/m\varepsilon(1+q^2d^2/\alpha^2)$, is plotted (for $\alpha=\sqrt{3}$) in the solid green line in Fig.~\ref{fig:system}. It is evident that the given model~(\ref{Eq:model}) describes the exact spectrum with sufficient accuracy up to the wavevector values of $qd\approx 3$. Therefore, $G_m$ provides a much better approximation than $G_e$, as it is applicable over a wider range of $qd \lesssim 1$ compared to the traditional expansion approach valid only for $qd \ll 1$.

It is worth mentioning that overall, considering $G_m$ instead of $G_e$ is analogous to using the regularized rather than ordinary Korteweg--De Vries equation~\cite{Benjamin1972}. In hydrodynamics, for example, a similar simplification of the integral equations has been successfully applied to describe non-linear waves and solitons on the surface of a liquid, including their peaking and breaking~\cite{WhithamBook}.  

In 2D plasma physics, a similar method of replacing the kernel in the Poisson equation was employed, most likely for the first time, to analyze the (linear) edge plasma oscillations in a half-plane 2DES~\cite{Fetter1985}. Afterwards, it was widely used to describe plasmons in various bounded or inhomogeneous 2DESs, including conventional systems~\cite{Cataudella1987,Mikhailov1995,Zabolotnykh2016,Cohen2018} with possible anisotropy~\cite{Stauber2019,Sokolik2021}, graphene~\cite{Wang2012,Hasdeo2017}, and topological insulators~\cite{Song2016,Kumar2016}.

Now let us proceed to the analytical description of a 2D plasma soliton. At this point, besides the Poisson equation~(\ref{Eq:Po}), we need to consider additional equations characterizing electron dynamics in a 2DES.
In this Letter, we follow a standard approach of using the Euler equation for the drift velocity of electrons $v(x,t)$ and the continuity equation for the deviation in the electron concentration $n(x,t)$ from its equilibrium value $n_0$:
\begin{equation}\label{Eq:basic}
\begin{split}
     &\partial_t v(x,t) +\partial_x \frac{v^2(x,t)}{2}=\frac{e}{m}\partial_x\varphi(x,t), \\
     &\partial_t n(x,t) +\partial_x \left[ (n_0+n(x,t))v(x,t)\right]=0.
\end{split}
\end{equation}

\begin{figure*}
    \includegraphics[width=2.0\columnwidth]{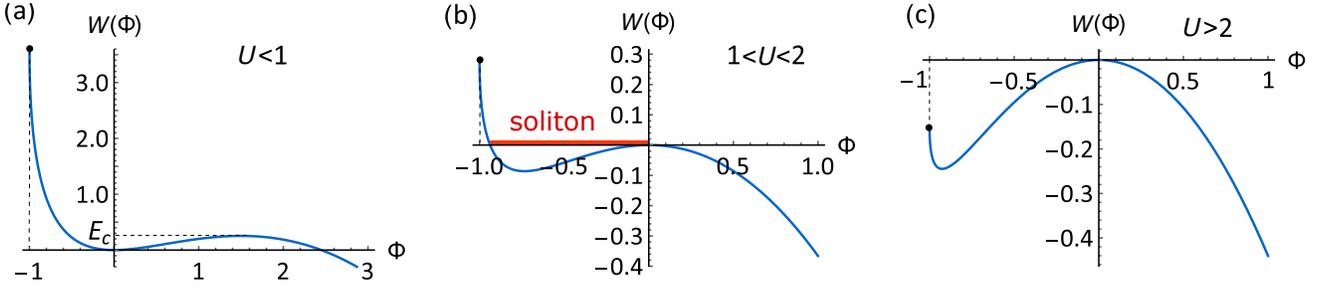}
    \caption{\label{fig:W} The effective potential energy $W(\Phi)$~(\ref{Eq:W}) plotted for different values of the dimensionless soliton velocity~(\ref{Eq:dim}) $U=0.7$ (a), $U=1.6$ (b), and $U=2.4$ (c). Solitons correspond to the finite solutions with zero energy that exist only for $1<U<2$, as indicated by red line in plot~(b).}
\end{figure*}

We seek the solutions in the form of a wave traveling along the $x$-axis with the velocity $u$. Hence, we introduce the argument $\xi=x-ut$. After this substitution, Eqs.~(\ref{Eq:basic}) can be integrated. Then, considering the result along with Eq.~(\ref{Eq:Po}), we obtain the following set of equations:
\begin{equation}\label{Eq:mod}
\left\{ 
\begin{split}
     & -u v(\xi) +\frac{v^2(\xi)}{2}-\frac{e}{m}\varphi(\xi)=0, \\
     & -u n(\xi) +\left[n_0+n(\xi)\right]v(\xi)=0, \\
     &\varphi(\xi)=\frac{-e}{\varepsilon}\int^{+\infty}_{-\infty}G(\xi-\xi')n(\xi')d\xi',
\end{split}
\right.
\end{equation}
where in the first two equations, the integration constants are set equal to zero since we seek the solitons with $v$, $n$, and $\varphi$ vanishing as $\xi \to \pm \infty$.

From the second equation in~(\ref{Eq:mod}), the velocity can be expressed as 
\begin{equation}
\label{Eq:v}
    v(\xi)= \frac{u n(\xi)}{n_0+n(\xi)}.
\end{equation}
Here, the denominator cannot equal zero or change its sign as the deviation $-n(\xi)$ cannot exceed $n_0$.

Next, we substitute $v(\xi)$ from (\ref{Eq:v}) into the first equation in (\ref{Eq:mod}) and, subsequently, $n(\xi)$ into the Poisson equation. Thus, we arrive at a single expression for $\varphi(x,t)$ that can be rewritten in terms of the dimensionless soliton potential $\Phi(\xi)=2e\varphi(\xi)/mu^2$ as follows:
\begin{equation}
\label{Eq:int}
    \Phi(\xi)=\frac{2 e^2 n_0}{m \varepsilon u^2}\int^{+\infty}_{-\infty}G(\xi-\xi')\left(1-\frac{1}{\sqrt{1+\Phi(\xi')}}\right)d\xi'
\end{equation}
At the same time, the condition of $n_0+n(\xi)>0$ implies the inequality $\Phi(\xi)>-1$. 

The derived non-linear integral equation~(\ref{Eq:int}) fully describes solitons in a gated 2DES. In principle, such and similar equations can be solved numerically~\cite{Narahara2008,Vostrikova2007,Liu2007,Suzuki2008,Zhiyenbayev2019}. Nevertheless, in the following discussion, we focus on the analytical treatment of Eq.~(\ref{Eq:int}) to achieve a comprehensible and straightforward description of solitons. 

As was mentioned above, to find the analytical solution, we substitute $G_m(\xi)$ from (\ref{Eq:model}) for $G(\xi)$ in (\ref{Eq:int}), which greatly simplifies Eq.~(\ref{Eq:int}) since $G_m(\xi)$ is essentially the Green's function of the differential operator 
\begin{equation}
\label{Eq:L0}
    \widehat L_m=\frac{d}{2 \pi \alpha^2} \left(- \partial^2_{\xi} +\frac{\alpha^2}{d^2}\right).
\end{equation}

Hence, applying $\widehat L_m$ to Eq.~(\ref{Eq:int}) after the substitution yields the following differential equation:
\begin{equation}
\label{Eq:osc}
    \Phi''(X)-\alpha^2 \left(\Phi(X)+\frac{2}{U^2}\left[\frac{1}{\sqrt{1+\Phi(X)}}-1\right] \right)=0.
\end{equation}
Here we introduce the dimensionless argument $X$ and the soliton velocity $U$ defined as:
\begin{equation}
\label{Eq:dim}
    X=\frac{\xi}{d}, \quad U=\frac{u}{v_p},\quad \text{where} \quad v_p=\sqrt{\frac{2\pi e^2 n_0 d}{m \varepsilon}}.
\end{equation}
In this case, $v_p$ is the velocity of gated plasmons (\ref{Eq:spectrum}) considered in the long-wavelength limit $qd \ll 1$~\cite{Chaplik1972}, and the double prime in Eq.~(\ref{Eq:osc}) denotes the second derivative with respect to $X$.
We note that Eq.~(\ref{Eq:osc}) in effect represents a non-linear oscillator defined as $\Phi''+V(\Phi)=0$. Naturally, it can be integrated to obtain an equivalent of the energy conservation law. Thus, multiplying Eq.~(\ref{Eq:osc}) by $\Phi'(X)$ and then integrating it results in the following relation:
\begin{eqnarray}
     & \frac{\bigl(\Phi'(X)\bigr)^2}{2}+\alpha^2 W(\Phi)=E, \quad \text{where} \label{Eq:En}\\
     & W(\Phi)=-\frac{\Phi^2}{2}+\frac{2}{U^2}\left(2+\Phi-2\sqrt{1+\Phi}\right)\label{Eq:W}.
\end{eqnarray}
Here, the ''total energy'' constant $E$ is chosen so that $W(0)=0$. 
It is important to emphasize that the model parameter $\alpha$ is included in the ''potential energy'' term $\alpha^2 W(\Phi)$ simply as a scaling factor. The function $W(\Phi)$ itself and the features of the potential energy are entirely independent of $\alpha$. In fact, the dimensionless soliton velocity $U$ is the only critical factor that defines the shape of $W(\Phi)$. Therefore, it is natural to analyze the characteristic patterns of $W(\Phi)$ for different values of $U$, as indicated in Fig.~\ref{fig:W}.

For solitons, $\Phi(X)$ and $\Phi'(X)$ tend to zero at $X\to \pm\infty$. Consequently, both terms in the left-hand side of Eq.~(\ref{Eq:En}) vanish, which, in turn, leads to $E=0$.
From Fig.~\ref{fig:W} it is clear that finite solutions $\Phi(X)$ with $E=0$ exist only when the velocity satisfies the condition of $1<U<2$. 
Although the investigation of the entire structure of non-linear waves is beyond the scope of this Letter, for a more consistent presentation, we provide a brief qualitative analysis of the cases for $U<1$ and $U>2$ as follows.
When $U<1$, there are finite solutions with $E>0$, where $\Phi(X)$ oscillates about zero value. These solutions most likely correspond to non-linear plasma oscillations. In the limit of small amplitude (when $E \to 0$), they take the standard form with the spectrum in~(\ref{Eq:spectrum}). As the amplitude and $E$ increase (assuming that $U$ remains constant), when $E$ becomes larger than $E_c$ (Fig.~\ref{fig:W}(a)), the solution ceases to be finite, and the oscillations break down, which is ordinary for non-linear oscillations. However, it should be noted that in the derivation of Eqs.~(\ref{Eq:En}) and (\ref{Eq:W}), we rely on the condition of $n$, $\varphi$, and $v$ vanishing at $\xi\to \infty$, which is not satisfied for the oscillatory regime. Therefore, the analysis above is only qualitative.
For $U>2$, we have $W(\Phi=-1)<0$, and all waves with $E$ near zero are likely to be unstable, breaking up to form shock waves, similar to the case of ion-acoustic waves in gaseous plasma~\cite{Sagdeev1966}.

\begin{figure}
    \includegraphics[width=1.0\columnwidth]{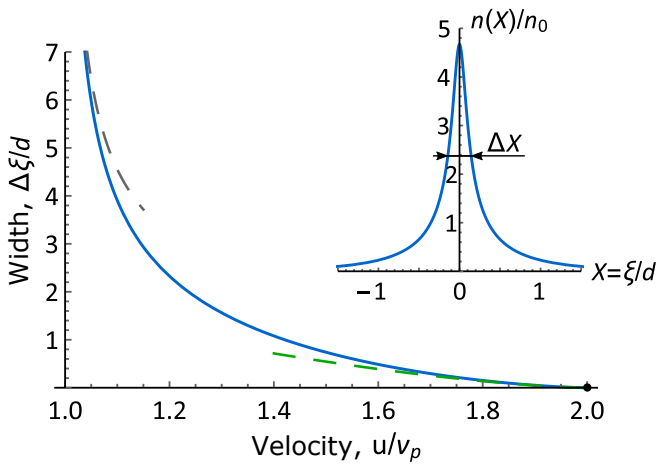}
    \caption{\label{fig:width} The dimensionless soliton width $\Delta X=\Delta\xi/d$ extracted from the concentration dependency $n(X)$, plotted as a function of the velocity $U=u/v_p$~(\ref{Eq:dim}). Gray and green dashed lines denote the asymptotics~(\ref{Eq:largeW}) and (\ref{Eq:smallW}), correspondingly. The width vanishes at $u/v_p=2$. The inset shows the soliton shape at $u/v_p=1.7$.}
\end{figure}

Now, let us expound on the solitonic regime of $1<U<2$ in particular. In this case, the amplitude of the soliton potential $\Phi_{max}$ and the concentration $n_{max}$ can be determined on account of the vanishing potential energy. Hence, $W(\Phi_{max})=0$ leads to $-\Phi_{max}=4(U-1)/U^2$ and
\begin{equation}
    \frac{n_{max}}{n_0}=\frac{1}{\sqrt{1+\Phi_{max}}} -1=2\cdot\frac{U-1}{2-U}.
\end{equation}
More significantly, Eqs.~(\ref{Eq:En}) and (\ref{Eq:W}) can be integrated exactly to provide an explicit relationship between the potential $\Phi$ (along with the concentration $n$) and $X=(x-ut)/d$. As the resulting expression is rather cumbersome, it is included in the Supplementary Material \footnote{See Supplementary Material below for the integrated relation between the potential $\Phi$ and $X=(x-ut)/d$.}. The obtained equation allows us to calculate the soliton width as a function of velocity $U$, as shown in Fig.~\ref{fig:width}. 

We consider the two limiting cases to find a straightforward analytical expression for the soliton width extracted from the concentration dependency $n(X)$.
First, in the low-velocity regime of $0<U-1\ll 1$, corresponding to the limit of weak non-linearity, the full soliton width $\Delta X$ at the half $n_{max}$ can be defined as
\begin{equation}
\label{Eq:largeW}
    \Delta X=\frac{\Delta\xi}{d}=\frac{2\sqrt{2}\ln(1+\sqrt{2})}{\alpha\sqrt{U-1}},
\end{equation}
which is in agreement with the previous studies of Korteweg--De Vries-like plasma solitons~\cite{Govorov1999}.

In the second limiting case of $0<2-U\ll 1$, the full width can be expressed as
\begin{equation}
\label{Eq:smallW}
     \Delta X=\frac{\Delta\xi}{d}=\frac{8}{3\alpha} \cdot (2-U)^{3/2}.
\end{equation}

Clearly, it follows from Eq.~(\ref{Eq:smallW}) that the width $\Delta X$ tends to zero as $U \to 2$. In this case, it invalidates the given approach as it violates the assumption for the parameter $qd\approx d/\Delta\xi=(\Delta X)^{-1}$ to be of the order of unity or less required for valid substitution of $G_m$ for $G$. Therefore, we cannot conclude with certainty whether solitons with $U\gtrsim 2$ exist.

In the above analysis we assume that the electron relaxation time $\tau$ is infinitely large. The finite value of $\tau$ in the framework of our approach leads to the appearance of an additional term $v(x,t)/\tau$ in the left-hand side of the first equation in (\ref{Eq:basic}). That term leads to the damping of solitons (as well as usual plasma waves) over the length $L$, which can be estimated as $L=u\cdot \tau$. Clearly, as the soliton propagates, its amplitude decreases and its shape is not preserved. Therefore, in that sense, solitons do not exist at finite values of $\tau$ (see also discussion after Eq.~(11) in Ref.~\cite{Narahara2008}).

Last but not least, let us stress the significance of considering a \textit{gated} 2DES. The critical issue is that for a soliton to be stable, its velocity must not coincide with the phase velocity of a linear wave. Otherwise, the soliton undergoes the decay with emission of linear waves~\cite{Zakharov1998}. In an ungated 2DES, linear plasmons have a square-root dispersion law $\omega\propto \sqrt{q}$~\cite{Stern1967} (corresponding to the limit of $qd\gg 1$ in the spectrum (\ref{Eq:spectrum})), i.e., their phase velocities take on any value. Hence, stable solitons cannot exist in such a system (see also discussion in Sec.~VI.B in Ref.~\cite{Zdyrski2019}). However, the presence of the gates (or, at least, one gate) limits the range of possible velocities of linear plasmons by $v_p$~(\ref{Eq:dim}), allowing for the solitons with velocities of $u>v_p$.

In summary, we consider solitary plasma waves in a 2DES with ideal metallic gates in its vicinity. In our investigation, we employ a non-perturbative model~(\ref{Eq:model}) that permits finding exactly and analytically the soliton velocity, amplitude, and shape. We overcome the drawback of conventional perturbative methods restricted by weak non-linearity conditions. In contrast, our technique enables describing the regime beyond weak non-linearity and dispersion limitations. In other words, it applies when the concentration deviation is of the order of the equilibrium concentration and the soliton width is comparable to the separation distance between the 2DES and the gate. We establish that solitons exist provided that their velocity $u$ lies in a finite range. We find the lower limit of $u$ to be determined by the maximum velocity of linear plasmons $v_p$~(\ref{Eq:dim}). As $u$ reaches the higher limit $2 v_p$, the soliton width (based on the concentration dependency) approaches zero, while the amplitude tends to infinity. We conclude that solitons do not exist beyond the specified range, for $u>2v_p$. Although we consider an ordinary 2DES with simple dynamics and constant effective mass of charges~(\ref{Eq:basic}), we believe that the proposed approach can be useful in studies of non-linear waves in various advanced structures, for example, in van der Waals heterostructures, in which charge dynamics is much more complex.

\begin{acknowledgments}
The author is grateful to Igor Zagorodnev and Vladimir Volkov for valuable discussions.
The work was financially supported by the Russian Science Foundation (Project No. 21-72-00114).
\end{acknowledgments}

\newpage
\onecolumngrid
\appendix

\section{\large{Supplementary Material for\\ ``Plasma solitons in gated two-dimensional electron systems: exactly solvable analytical model for the regime beyond weak non-linearity''}} 
To derive the relationship between the potential $\Phi$ and $X=(x-ut)/d$, we can rewrite Eqs.~(12) and (13) from the main text in the following form:
\begin{equation}
\label{eq1}
    \int^{\Phi}_{\Phi_{max}}\frac{d\Phi}{\sqrt{-\frac{2}{U^2}\left(2+\Phi- 2\sqrt{\Phi+1}\right)+\Phi^2/2}}=\alpha\sqrt{2}\int^{X}_0 dX,
\end{equation}
where $\alpha$ is the model constant equal $\sqrt{3}$~\cite{CarrierBook}, and the integration limits are chosen so that $\Phi(X=0)=\Phi_{max}$.
Considering $\sqrt{\Phi+1}$ as a new function, the integral on the left-hand side of Eq.~(\ref{eq1}) can be found exactly. Hence, Eq.~(\ref{eq1}) can be written in a different form as follows:
\begin{equation}
\label{eq3}
\begin{split}
    &-2\ln \left(\frac{U S_1}{2}\right)+
    \left(1-U^{-2}\right)^{-1}\ln\left(US_2\right)=\alpha X, \quad \text{where}  \\
    & S_1=\sqrt{\Phi+1}+1+\sqrt{\left(\sqrt{\Phi+1}+1\right)^2-4U^{-2}} \quad \text{and}\\
    & S_2=\sqrt{(1-U^{-2})\left(4\cdot\frac{1-U^{-2}}{(1-\sqrt{\Phi+1})^2} -\frac{4}{1-\sqrt{\Phi+1}}+1\right)}+2\frac{1-U^{-2}}{1-\sqrt{\Phi+1}}-1.
\end{split}    
\end{equation}
Thus, we have determined the direct relation between $\Phi$ and $X$.
To find the deviation in the concentration $n(X)$, we can substitute into Eq.~(\ref{eq3}) the following relation between $\Phi$ and $n$:
\begin{equation}
    \frac{n(X)}{n_0}=\frac{1}{\sqrt{1+\Phi(X)}}-1,
\end{equation}
where $n_0$ is the equilibrium concentration.

\end{document}